\documentclass[prx,superscriptaddress,twocolumn,aps]{revtex4}
\usepackage{amssymb}
\usepackage{graphicx}
\usepackage{longtable}
\usepackage{amsmath}
\usepackage{amsfonts}
\usepackage{color}
\begin{document}

\author{A. Chiesa}
\affiliation{Dipartimento di Fisica e Scienze della Terra, Universit\`a di Parma, I-43124 Parma, Italy}
\author{P. Santini}
\affiliation{Dipartimento di Fisica e Scienze della Terra, Universit\`a di Parma, I-43124 Parma, Italy}
\author{D. Gerace}
\affiliation{Dipartimento di Fisica, Universit\`a di Pavia, via Bassi 6, I-27100 Pavia, Italy}
\author{J. Raftery}
\affiliation{Department of Electrical Engineering, Princeton University, Princeton, New Jersey 08544, USA}
\author{A. A. Houck}
\affiliation{Department of Electrical Engineering, Princeton University, Princeton, New Jersey 08544, USA}
\author{S. Carretta}
\affiliation{Dipartimento di Fisica e Scienze della Terra, Universit\`a di Parma, I-43124 Parma, Italy}

\date{\today }
\title{ Digital quantum simulators in a scalable architecture of hybrid spin-photon qubits }
\begin{abstract}
Resolving quantum many-body problems represents one of the greatest challenges in physics and physical chemistry, due to the prohibitively large computational resources that would be required by using classical computers. A solution has been foreseen by directly simulating the time evolution through sequences of quantum gates applied to arrays of qubits, i.e. by implementing a \textit{digital quantum simulator}. Superconducting circuits and resonators are emerging as an extremely-promising platform for quantum computation architectures, but a digital quantum simulator proposal that is straightforwardly scalable, universal, and realizable with state-of-the-art technology is presently lacking.
Here we propose a viable scheme to implement a universal quantum simulator with hybrid spin-photon qubits in an array of superconducting resonators, which is intrinsically scalable and allows for local control. As representative examples we consider the transverse-field Ising model, a spin-1 Hamiltonian, and the two-dimensional Hubbard model; for these, we numerically simulate the scheme by including the main sources of decoherence. In addition, we show how to circumvent the potentially harmful effects of inhomogeneous broadening of the spin systems.
\end{abstract}
\maketitle

\vskip 3 cm

\section{Introduction}
There is a large number of problems that are well known to be hardly tractable with standard computational approaches and resources, mainly due to the many-body nature of strongly correlated many particle systems. To overcome this limitation, the idea of a quantum simulator was originally proposed by Feynman \cite{Feynman}: any arbitrary complex quantum systems could in fact be simulated by another quantum system mimicking its dynamical evolution, but under the experimenter control. This idea was later refined and mathematically formalized in quantum information perspectives by Lloyd \cite{Lloyd}.\\
Over the past twenty years, different approaches have been proposed to realize quantum simulators of the most relevant models in condensed matter physics, quantum field theories, and quantum chemistry \cite{Norisim}. Most efficient protocols have been proposed  and experimentally realized with trapped ions \cite{Lamata_review, Blatt}.
Generally speaking, quantum simulators can be broadly classified into two main categories: in \textit{digital simulators} the state of the target system is encoded in qubits and its Trotter-decomposed time evolution is implemented by a sequence of elementary quantum gates \cite{Lloyd}, whereas in \textit{analog simulators} a certain quantum system directly emulates another one. Digital architectures are usually able to simulate broad classes of Hamiltonians, whereas analog ones are restricted to specific target Hamiltonians. For a recent review on these different approaches, we refer to \cite{Norisim} and references therein.\\
Lately, superconducting circuits and resonators have emerged as an extremely promising platform for quantum information and quantum simulation architectures \cite{KochRev,HouckSim,PRLSolano,arxivSolano}. The first and unique proposal for a general-purpose digital simulator has been put forward only very recently \cite{PRLSolano}. In this proposal qubits encoded in transmons are dispersively coupled through a photon mode of a single resonator, and such coupling is externally tuned by controlling the transmon energies. However, the reported fidelities and the intrinsic serial nature of this setup (i.e., the need of addressing each pair of qubits sequentially), may hinder the scalability to a sizeable number of qubits. 
In addition, superconducting units are not ideal for encoding qubits owing to their relatively short coherence times. Indeed, spin-ensembles \cite{KuboPRL,PRBRspins,EchoBertet} or even photons
\cite{Mariantoni,Ghosh} have been proposed as memories to temporarily store the state of superconducting computational qubits.\\
Here we consider an array of superconducting resonators as the main
technological platform, on which hybrid spin-photon qubits are defined by introducing strongly coupled spin ensembles (SEs) in each resonator \cite{PRLcav,PRAcav}. One- and two-qubit quantum gates can be implemented by individually and independently tuning the resonators modes through external magnetic fields.  This setup can realize a universal digital quantum simulator, whose scalability to an arbitrary large array is naturally fulfilled by the inherent definition of the single qubits, represented by each coupled SE-resonator device. The possibility to perform a large number of two-qubit gates in parallel makes the manipulation of such large arrays much faster than in a serial implementation, thus making the simulation of complex target Hamiltonians possible in practice.\\
A key novelty of the present proposal is that ensembles of effective $S=1$ spins are used in the hybrid encoding, which allows to exploit the mobility of photons across different resonators to perform two-qubit gates between physically distant qubits. This is done much more efficiently than by the straightforward approach of moving the states of the two qubits close to each other by sequences of SWAP gates, and makes the class of Hamiltonians which can be realistically addressed much larger. Long-distance operations arise whenever mapping the target system of the simulation onto the register implies two-body terms between distant qubits. Besides the obvious case of Hamiltonians with long-range interactions, this occurs with any two-dimensional model mapped onto a linear register, or with models containing $N$-body terms, including the many-spin terms which implement the antisymmetric nature of fermion wavefunctions.\\
The time evolution of a generic Hamiltonian is decomposed into a sequence of local unitary operators, which can be implemented by means of elementary single- and two-qubits gates. Then we combine the elementary gates of our setup in order to mimic the dynamics of spin and Hubbard-like Hamiltonians for fermions.
We explicitly report our results for the digital quantum simulation of the transverse-field Ising model on 3 qubits, the tunneling dynamics of a spin one in a rhombic crystal field and the Hubbard Hamiltonian. The robustness of the scheme is demonstrated by including the effects of decoherence in a master equation formalism.
Finally, we discuss the main sources of errors in the present simulations and the possibility to overcome them. In particular, we show how potentially harmful effects of inhomogeneous broadening of the spin ensemble are circumvented by operating the scheme in a cavity-protected regime.

\begin{figure}
   \centering
   \includegraphics[width=8cm]{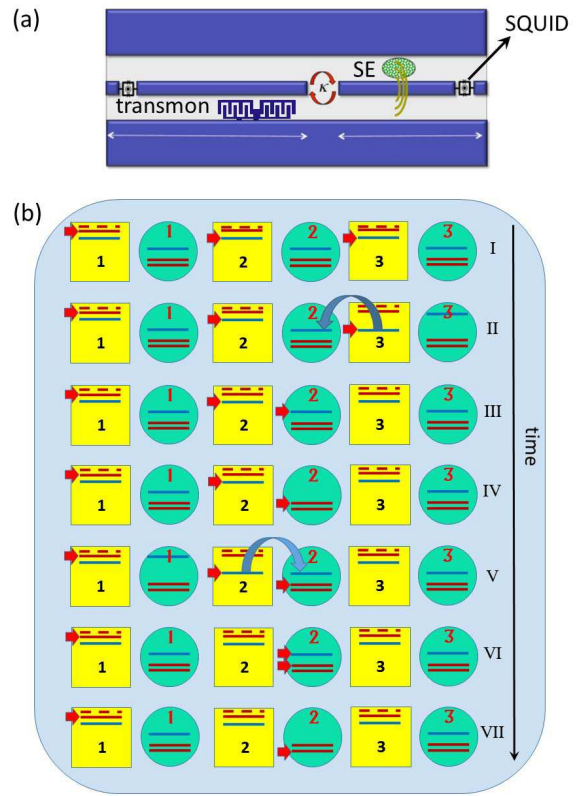}
   \caption{(a) Elementary unit of the scalable setup, consisting of an {\it auxiliary} and a {\it logical} resonator. The latter includes an ensemble of $S=1$ spins, placed at the antinodes of the magnetic field (rotational lines) of the cavity mode. The {\it auxiliary} resonator contains a nonlinear element (transmon) coupled to the electric field of the fundamental mode. (b) Detailed sequence of time steps required to produce controlled-$\varphi$ two-qubit gate between qubits $\mu = 2$ and $\mu=3$ (see Appendix B for details). {\it Logical} cavities are represented by square boxes, whereas {\it auxiliary} resonators are depicted as circular boxes. Blue lines represent photon frequencies in the idle configuration ($\omega_c^\mu(0)$ in the {\it logical} and $\tilde{\omega}_c^j(0)$ in the {\it auxiliary} cavities). The transmon ($\Omega_{01}$ and $\Omega_{12}$) and spin ($\omega_{-1}$, continuous, and $\omega_{1}$, dashed) transition energies are indicated by red lines.
   (I) qubits are initially into state $|1_2 1_3\rangle$, with the excitations (red arrows) stored into the photonic degrees of freedom (blue lines); (II) {\it logical} cavity 3 is brought into resonance with the {\it auxiliary} resonator $j=2$, thus (III) bringing the photon to the auxiliary cavity. In the meantime {\it auxiliary} resonator 3 is detuned from the others to avoid unwanted photon hoppings. In (IV) the photon is absorbed by the transmon ($|\psi_{0,j=2}\rangle \rightarrow |\psi_{1,j=2}\rangle$ transition). The same hopping process (V) is repeated for the photon originally in cavity 2, which is brought to the {\it auxiliary} resonator (VI) and then absorbed and emitted by the transmon ($|\psi_{1,2}\rangle \rightarrow |\psi_{2,2}\rangle$ transition) in a semi-resonant Rabi process (VII). The procedure is then repeated to bring photons back to {\it logical} cavities 2 and 3, leading the state back to $|1_2 1_3\rangle$ with an additional phase $\varphi$ acquired during the semi-resonant process.}
   \label{Setup}
\end{figure}



\section{A scalable architecture for quantum simulation}
The proposed quantum simulator is schematically shown in Fig.~\ref{Setup}. It consists of a one- or two-dimensional (1D or 2D) lattice of superconducting resonators where hybrid spin-photon qubits are defined. We notice that large arrays of such resonators have already been shown experimentally~\cite{Underwood2012,HouckSim}.
In this schematic implementation, qubits are encoded within square boxes. Each box represents a coplanar resonator containing an ensemble of (effective) $S=1$ spins, whose collective excitations correspond to the transitions from the $m = 0$ single-spin ground state to the $m = \pm 1$ excited states, and can be modeled by two independent harmonic oscillators.
Red lines represent the transition energies (continuous $m=-1$, dashed $m=1$ transitions, respectively), while the blue line indicates the resonator frequency in the idle configuration. This can be varied within a nanosecond time-scale by means of SQUID devices properly connected to the resonator \cite{tunability1,tunability2,tunabilityns}, in order to match the spin transition frequencies.
In the hybrid qubit encoding, a dual-rail representation of the logical units is introduced where the $| 0 \rangle_\mu$ and $| 1 \rangle_\mu$ states of qubit $\mu$ are defined in the single-excitation subspace of each resonator. The logical state $| 0 \rangle_\mu$ ($| 1 \rangle_\mu$) corresponds to zero (one) photons and a single (zero) quantum in the $m=-1$ oscillator in cavity $\mu$. This encoding has been introduced in previous works~\cite{PRLcav,PRAcav}, and it is detailed in App.~A for completeness. The $m=1$ oscillator represents an auxiliary degree of freedom that is exploited to store the photonic component of the qubit, if needed (e.g., to perform two-qubit gates between distant qubits, see App.~B).\\
The basic unit of the scalable array is represented by a pair of qubits connected by an interposed auxiliary resonator containing a superconducting transmon device (circular box), which is employed to perform two-qubit gates. It should be emphasized that this nonlinear superconducting element is not used to encode any information, and {\it it is left in its ground state always except during the implementation of the two-qubit gates}. Consequently, its possibly short coherence times do not significantly affect the quantum simulation.\\
In the following, we shall refer to the square boxes as the {\it logical} cavities labelled with Greek letters, while the circular ones are the {\it auxiliary} cavities labeled by Latin letters.
Photon hopping between neighboring resonators is allowed by capacitive coupling. Formally, such a complex system can be described by the total Hamiltonian
\begin{equation}
\hat{H} = \hat{H}_{spin} + \hat{H}_{tr} + \hat{H}_{ph}   + \hat{H}_{int} + \hat{H}_{ph-ph} \, .
\label{hamil}
\end{equation}
The first term describes the SEs as independent harmonic oscillators \cite{KuboPRA} ($\hbar \equiv 1$):
\begin{equation}
\hat{H}_{spin} =  \sum_m \sum_\mu \omega_m \hat{b}_{m,\mu}^{\dagger} \hat{b}_{m,\mu} \, ,
\label{hamil2}
\end{equation}
where $\hat{b}^\dagger_{m,\mu}$ creates a spin excitation in level $m=\pm 1$ of resonator $\mu$.
The transmons are treated as effective three-level systems, with transition energies $\Omega_{01}$ and $\Omega_{12}$, and described by
\begin{equation}
\hat{H}_{tr} =  \sum_j \Omega_{01} |\psi_{1,j} \rangle \langle \psi_{1,j}| + (\Omega_{12}+\Omega_{01}) |\psi_{2,j} \rangle \langle \psi_{2,j}| .
\label{HamCPB}
\end{equation}
The time-dependent photonic term is entirely responsible for the manipulation of the qubits. It can be expressed as:
\begin{equation}
\hat{H}_{ph} = \sum_{\mu} \omega_c^\mu (t) \hat{a}^{\dagger}_\mu \hat{a}_\mu +  \sum_{j} \tilde{\omega}_c^j (t) \hat{\tilde{a}}^{\dagger}_j \hat{\tilde{a}}_j \, ,
\label{hamil3}
\end{equation}
where $ \omega_c^\mu (t) = \omega_c^\mu (0) + \delta_c^\mu (t) $ and a similar expression holds for $\tilde{\omega}_c^j(t)$. $\hat{a}^{\dagger}_\mu$ ($\hat{a}_\mu$) creates (destroys) a single photon in the {\it logical} resonator $\mu$, while $\hat{\tilde{a}}^{\dagger}_j$ ($\hat{\tilde{a}}_j$) creates (destroys) a single photon in the {\it auxiliary} cavity $j$.
Hereafter, we will use the interaction picture, with $ \hat{H}_0 = \hat{H}_{spin}+ \hat{H}_{tr} +\hat{H}_{ph}(t=0)$.
Hence, within the rotating-wave approximation the spin-photon and transmon-photon coupling Hamiltonian takes the form:
\begin{eqnarray} \nonumber
\hat{H}_{int}  &=&
\!\!G_{01}\!  \sum_j \left[ \hat{\tilde{a}}^{\dagger}_j |\psi_{0,j}\rangle \langle \psi_{1,j} | e^{i(\tilde{\omega}_c^j-\Omega_{01}) t} + \text{h.c.} \right] \\ &+& \nonumber
\!\!G_{12}\! \sum_j \left[ \hat{\tilde{a}}^{\dagger}_j |\psi_{1,j} \rangle \langle \psi_{2,j} | e^{i(\tilde{\omega}_c^j-\Omega_{12}) t} + \text{h.c.} \right] \\ &+&
\!\!\! \sum_{m=1,-1} \sum_\mu \bar{G}_{m}\! \left [ \hat{a}^{\dagger}_\mu \hat{b}_{m,\mu} e^{i(\omega_c^\mu-\omega_m) t} + \text{h.c.} \right].
\label{hamil4}
\end{eqnarray}
Here, the coupling constants $\bar{G}_{m}$ for the SE are enhanced with respect to their single-spin counterparts by a factor $\sqrt{\mathcal{N}}$, $\mathcal{N}$ being the number of spins in the SE \cite{Wesenberg}.\\
Finally, the last term in  Eq.~(\ref{hamil}) describes the photon-hopping processes induced by the capacitive coupling between the modes in neighboring cavities \cite{Underwood2012}:
\begin{equation}
\hat{H}_{ph-ph} = - \kappa \sum_{\langle \mu,j \rangle}  \hat{a}^{\dagger}_\mu \hat{\tilde{a}}_j e^{i(\omega_c^\mu-\tilde{\omega}_c^j) t} + \text{h.c.} \,.
\label{hopping}
\end{equation}
Single- and two-qubit gates are efficiently implemented by tuning individual resonator modes, as shown in previous works \cite{PRLcav,PRAcav}. Arbitrary single-qubit rotations within the Bloch sphere as well as controlled-phase (C$\varphi$) gates can be realized (see App.~B for a summary).\\
The present setup offers two remarkable benefits: the first is that using the hybrid encoding with an ensemble of effective $S=1$ spins ensures the possibility of implementing Controlled-phase gates between distant qubits, with no need of performing highly demanding and error-prone sequences of SWAP gates.
This is done by bringing the photon components of the two qubits into neighboring {\it logical} resonators by a series of hopping processes (see App.~B for details).
Transferring the photons with no corruption and without perturbing the qubits encoded in the interposed {\it logical} cavities is made possible by temporarily storing the photon component of these interposed qubits into the $m=1$ spin oscillator. \\
In addition, quantum simulations can be performed in parallel to a large degree, with resulting reduction of simulation times. This is made possible by the
definitions of the single qubits, represented by each coupled SE-resonator device, and by the local control of each {\it logical} or {\it auxiliary} resonator. Non-overlapping parts of the register can then be manipulated in parallel. For instance, in simulating a Heisenberg chain of $N$ spins $s=1/2$, the $N$ two-qubits evolutions which appear at each time-step in the Trotter decomposition are performed simultaneously first on all $N/2$ "even" bonds and then simultaneously on the remaining $N/2$ "odd" bonds. Thus the simulation time of each Trotter step does not increase with $N$.

\section{Numerical experiments}

While it is obvious that a universal quantum computer can be used in principle to simulate any Hamiltonian, the actual feasibility of such simulations needs to be quantitatively assessed by testing whether the complex sequences of gates needed are robust with respect to errors due to decoherence. Here we numerically solve the density matrix master equation for the model in Eq. (\ref{hamil}) with the inclusion of the main decoherence processes, i.e., photon loss and dephasing of the transmons \cite{PRAcav} (see App.~C for details). The potentially harmful effects of inhomogeneous broadening in the SE will be addressed in the next section.\\
In the following, we will consider the \textit{fidelity}
\begin{equation}
\mathcal{F} = \sqrt{\langle \psi | \hat{\rho} | \psi \rangle } \, ,
\label{Fidelity}
\end{equation}
as a valuable figure of merit for the target Hamiltonians to be simulated, where $\hat{\rho}$ is the final density matrix and $|\psi\rangle$ the target state.
For the simulations shown in the following, we have chosen these operational parameters: $\omega_1/2 \pi = 37$ GHz, $\omega_{-1}/2 \pi = 35$ GHz, $\omega_c(0)/2 \pi = 31$ GHz, $\tilde{\omega}_c(0)/2 \pi = 28$ GHz and $\Omega_{01}/2 \pi = 21.7$ GHz, $\Omega_{12}/2 \pi = 19.6$ GHz (see the level scheme inside each cavity in Fig. \ref{Setup}). We also assume realistic values of the SE-resonator $\bar{G}_{\pm 1} = 40 $ MHz, transmon-resonator $G_{01} = 30 $ MHz, $G_{12} = 40 $ MHz and photon-photon $\kappa = 30$ MHz couplings, respectively \cite{Schuster,Underwood2012}.
The transmon parameters correspond to a ratio between Josephson and charge energies $E_J/E_C = 25$ \cite{NoriRMP2013}. In this regime the dephasing time $T_2^{tr}$ exceeds several $\mu s$ while keeping a $10\%$ anharmonicity. The two chosen spin gaps can easily be achieved with several diluted magnetic ions possessing a $S>1/2$ ground multiplet, just by applying a small magnetic field along a properly chosen direction.
We have chosen resonator frequencies $\omega_c$ and $\tilde{\omega_c}$ larger than usual experiments (e.g., twice the typical frequencies reported in Ref.~\onlinecite{Schuster}), since this helps improving the maximal fidelity of gates. However, we emphasize that the results do not qualitatively depend on
these specific numbers. For instance, in the next Section we show that remarkably good fidelities can be obtained by using realistic and state-of-art experimental frequencies~\cite{Schuster} or smaller.

\subsection{Digital simulation of spin Hamiltonians}
\begin{figure}
   \includegraphics[width=8.5cm]{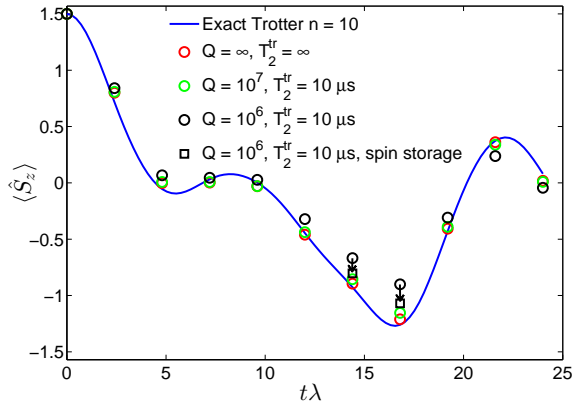}
   \caption{ Oscillations of the magnetization in the transverse-field Ising model. The simulation is performed on a chain of 3 qubits, in the case $b=\lambda/2$. The plot reports the expectation value of the total magnetization $\langle \hat{S}_z \rangle = $ tr$\left[\hat{\rho} (\hat{s}_{1z}+\hat{s}_{2z}+\hat{s}_{3z})\right]$ on the final state of the system, simulated for different values of the resonator quality factor (points) and compared with the exact evolution (line).}
   \label{Ising}
\end{figure}

Since most Hamiltonians of physical interest can be written as the sum of $L$ local terms, our quantum computing architecture can be employed to efficiently simulate the time-evolution induced by any target Hamiltonian of the type $\hat{\mathcal{H}} = \sum_k^L \hat{\mathcal{H}}_k $.
The system dynamics can be approximated by a sequence of unitary operators according to the Trotter-Suzuki formula ($\hbar=1$):
\begin{equation}
\hat{U}(t)=e^{-i \hat{\mathcal{H}} t} \approx ( e^{-i \hat{\mathcal{H}}_1 \tau} \cdots e^{-i \hat{\mathcal{H}}_L \tau} )^n ,
\label{Trotter}
\end{equation}
where $\tau = t/n$ and the total {\it digital} error of the present approximation can be made as small as desired by choosing $n$ sufficiently large \cite{Lloyd}.
In this way the simulation reduces to the sequential implementation of local unitaries, each one corresponding to a small time interval $t/n$.
The set of local unitary operators can be implemented by a proper sequence of single- and two-qubit gates.\\
The mapping of $s=1/2$ models onto an array of qubits is straightforward.
We consider here two kinds of significant local terms in the target Hamiltonian, namely
one- ($\hat{\mathcal{H}}^{(1)}_\alpha$) and two-body ($\hat{\mathcal{H}}^{(2)}_{\alpha \beta}$) terms, with $\alpha, \beta = x, y, z$.
The unitary time evolution corresponding to one-body terms $\hat{\mathcal{H}}^{(1)}_\alpha = b \hat{s}_\alpha$ is directly implemented by single-qubit rotations $\hat{R}_\alpha(b \tau)$.
Conversely, two-body terms describe a generic spin-spin interaction of the form $\hat{\mathcal{H}}^{(2)}_{\alpha \beta} = \lambda \hat{s}_{1 \alpha} \hat{s}_{2 \beta}$, for any choice of $\alpha, \beta = x, y, z$.
The evolution operator, $e^{-i \hat{\mathcal{H}}^{(2)}_{\alpha \beta} \tau }$, can be decomposed as \cite{PRLsimulatori}
\begin{equation}
e^{-i \lambda \hat{s}_{1 \alpha} \hat{s}_{2 \beta} \tau } = [\hat{u}_{1 \alpha} \otimes \hat{u}_{2 \beta}]\: e^{-i \hat{\Lambda} \tau}\: [\hat{u}_{1 \alpha} \otimes \hat{u}_{2 \beta}]^\dagger \, ,
\label{decomp}
\end{equation}
with $\hat{\Lambda} = \lambda \hat{s}_{1z} \hat{s}_{2z}$, $\hat{u}_x = \hat{R}_y (\pi/2) $, $\hat{u}_y = \hat{R}_x (3 \pi/2) $, $\hat{u}_z = \hat{I}$.
The Ising evolution operator, $e^{-i \lambda \hat{s}_{1z} \hat{s}_{2z} \tau}$, can be obtained starting from the two-qubit C$\varphi$ gate and exploiting the identity (apart from an overall phase)
\begin{equation}
e^{-i \lambda \hat{s}_{1z} \hat{s}_{2z} \tau} = [\hat{\Phi}_1(-\varphi/2) \otimes \hat{I}_2]\: \hat{U}_{C{\varphi}} \:[\hat{I}_1 \otimes \hat{\Phi}_2(-\varphi/2)],
\label{decompZZ}
\end{equation}
where $\varphi= \lambda \tau$. Here $\hat{\Phi}(\varphi)$ is a phase gate (see App.~B).
The time required and the fidelity for the simulation of each term of a generic spin Hamiltonian are calculated by using a Lindblad master equation formalism and are listed in Table \ref{Hamunits}.
We notice that the predicted fidelities are very high, even after the inclusion of realistic values for the main decoherence channels, especially for the photon loss rate $\Gamma_\mu$,
which is related to the resonators quality factor ($Q$) by $\Gamma_\mu = \omega_c^\mu/Q$.
These elementary steps can be used to simulate non-trivial multi-spin models.\\
\begin{table}
  \begin{tabular}{|c|c|c|c|c|}
    \hline
     & ~time~   & ~~~$\mathcal{F}_I$~~~  & ~~~$\mathcal{F}_D$~~~& ~~~$\mathcal{F}_D^{(CP)}$~~~   \\
    \hline
    $\mathcal{H}_x^{(1)}$ & 6.4 ~ns & 99.99~\% & 99.94~\% & 99.76~\% \\
    $\mathcal{H}_{z}^{(1)}$   & 0.5 ~ns & 99.99~\% & 99.98~\% & 99.86~\% \\
    $\mathcal{H}_{yy}^{(2)}$ & 85.8 ~ns & 99.87~\% & 99.24~\% & 98.69~\%  \\
    $\mathcal{H}_{zz}^{(2)}$ & 61 ~ns & 99.91~\% & 99.45~\%& 99.00~\%  \\
    $\mathcal{H}_{yz}^{(2)}$ & 85.8 ~ns & 99.79~\% &  99.13~\% & 98.57~\% \\
    \hline
  \end{tabular}
 \caption{{\bf Simulation of the elementary terms of the Hamiltonian.} Fidelity and time required to simulate the elementary terms of the Spin Hamiltonian. The fidelity has been calculated by assuming a random initial state. The second and third column show a comparison between the ideal fidelity (calculated in the absence of decoherence) and the real one (calculated assuming a Lindblad dynamics, with $Q=10^6$ and $T_2^{tr} = 10 ~\mu s$). The implemented evolution is $\hat{U} = \exp\left[-i \mathcal{H}_{\alpha \beta}^{(1,2)} \tau \right]$, with $b \tau = \lambda \tau = \pi/2$. The last column reports the fidelities with a setup operating in a cavity-protected regime, with $Q=10^6$ and $T_2^{tr} = 10 ~\mu s$ (see Section IV).}
  \label{Hamunits}
\end{table}
As a prototypical example we report the digital quantum simulation of the transverse field Ising model (TIM) on a chain of 3 qubits:
\begin{equation}
\hat{\mathcal{H}}_{TIM} = \lambda \left( \hat{s}_{1z} \hat{s}_{2z} + \hat{s}_{2z} \hat{s}_{3z} \right) + b \left( \hat{s}_{1x} + \hat{s}_{2x} + \hat{s}_{3x} \right) \, ,
\label{targetIsing}
\end{equation}
where $\hat{s}_{i \alpha}$ are spin-1/2 operators. Figure~\ref{Ising} shows the oscillations of the magnetization, Tr$\left[\hat{\rho} (\hat{s}_{1z}+\hat{s}_{2z}+\hat{s}_{3z})\right]$, for a spin system initialized in a ferromagnetic configuration. Here $\hat{\rho}$ is the three-qubit density matrix obtained at the end of the $n=10$ Trotter steps of the simulation. The exact Trotter evolution (continuous line) is compared to the simulated one (points). In particular, red circles represent the ideal evolution, without including any source of decoherence. Errors are, in that case, only due to a non-ideal implementation of the quantum gates (see discussion below). Conversely, green and black circles are calculated including the most important decoherence channels, namely photon loss (timescale $1/\Gamma_\mu$) and pure dephasing of the transmon (timescale $T_2^{tr}$). It turns out that photon loss is the most important environmental source of errors \cite{PRAcav}, while $T_2^{tr} \approx 10 ~\mu s$ \cite{deph} is sufficient to obtain high fidelities at the end of the simulation. Indeed, the transmon is only excited during the implementation of two-qubit gates.
The simulation has been performed for different values of the resonators quality factor. By decreasing $Q$ the average fidelity decreases from $96.5\%$ (infinite $Q$) to $94.6\%$ ($Q=10^7$) and $84.6 \%$ ($Q=10^6$). For high but realistic \cite{megrant} values of $Q = 10^7$ the calculated points are close to the ones obtained in the ideal case (with infinite $Q$): in that case the gating errors still dominate the dynamics.
Finally, by exploiting the auxiliary $m=1$ oscillator to store the photon component of the hybrid qubits when these are idle, the effects of photon loss are reduced and the fidelity significantly increases.
The improvement is apparent in Fig. \ref{Ising}, by comparing black circular and square points; the final fidelity raises from $84.6\%$ to $92\%$ thanks to this storage.
We stress again that the simulation time of each Trotter step does not increase for larger systems containing more than 3 spins.
Indeed, even if more gates are needed, these can be applied in parallel to the whole array, independently of the system size.

\begin{figure}
   \centering
  \includegraphics[width=8.5cm]{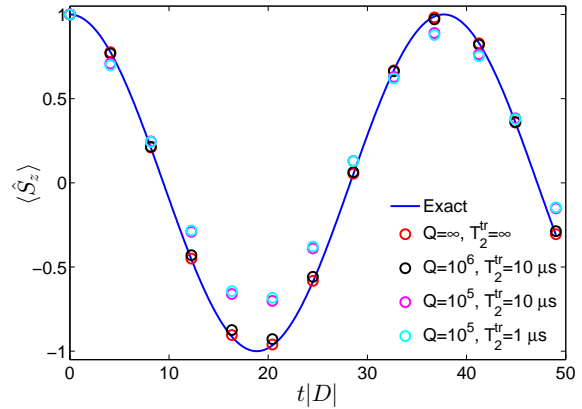}
   \caption{ Tunneling of the magnetization in a spin-1 system.
   Line: exact time evolution of $\langle \hat{S}_z\rangle$ for a single $S=1$ spin with $|D/E| = 12$, after Eq. (\ref{eqspin1tun}).
   As it is well known, the system oscillates between states with opposite magnetization 
   due to quantum tunneling across the anisotropy barrier.
   Points: digital quantum simulation obtained by the time evolution of two hybrid qubits for different values of the resonator quality factor,
   $Q$, and of the transmon coherence time, $T_2^{tr}$, respectively.}
   \label{spin1}
\end{figure}

The simulation of Hamiltonians involving $S > 1/2$ spin ensembles can be performed by encoding the state of each spin-$S$ onto that of $2 S$ qubits. As an explicit example, we consider a chain of $S=1$ spins, labelled $\hat{{\bf S}}_i$, with nearest-neighbor exchange interactions and single-spin crystal-field anisotropy, described by the Hamiltonian
\begin{equation}
\hat{\mathcal{H}}_{s1} = \sum_i \lambda  \hat{{\bf S}}_{i} \cdot \hat{{\bf S}}_{i+1}  + \sum_i \left[ D \hat{S}_{iz}^2 + E \left( \hat{S}_{ix}^2 - \hat{S}_{iy}^2\right) \right] \, ,
\nonumber
\end{equation}
which reduces to the paradigmatic Haldane case for $D=E=0$. By rewriting each spin-1 operator as the sum of two spin-1/2 ones ($\hat{S}_{i \alpha}=\hat{s}_{i A \alpha}+\hat{s}_{i B \alpha}$), $\hat{\mathcal{H}}_{s1}$ can be mapped onto a $s=1/2$ Hamiltonian, $\hat{\tilde{\mathcal{H}}}_{s1}$, with twice the number of spins. Indeed, if each A-B pair of qubits is initialized into a state with total spin equal to one, the dynamics of $\hat{\tilde{\mathcal{H}}}_{s1}$ coincides with that of $\hat{\mathcal{H}}_{s1}$ and can be simulated along the lines traced above.
A proof-of-principle experiment, which could be implemented even by the non-scalable single-resonator setup described in Ref.~\onlinecite{PRAcav}, would be the simulation of a single spin $S=1$ experiencing tunneling of the magnetization. In this simple case we find (apart from a constant term):
\begin{equation}
\hat{\tilde{\mathcal{H}}}_{s1} = 2 D \hat{s}_{zA} \hat{s}_{zB}  + 2 E \left( \hat{s}_{xA} \hat{s}_{xB} - \hat{s}_{yA} \hat{s}_{yB}\right).
\label{eqspin1tun}
\end{equation}
Figure~\ref{spin1} reports the comparison between the exact and the simulated evolution of the magnetization, assuming $D<0$ and $|D/E|=12$, for different values of $Q$ and $T_2^{tr}$. Interestingly, quantum oscillations of $\langle \hat{S}_z \rangle$ are well captured by the simulation even for $Q=10^5$, and the fidelity is practically unaffected by a reduction of transmon coherence time to $T_2^{tr}=1$ $\mu$s.\\
The simulation of many-spin models with $S>1$ typically requires two-qubit gates involving non-nearest-neighbor qubits. These can be handled with no need of SWAP gates as outlined in App.~B.

\subsection{Digital simulation of Fermi-Hubbard models}
The numerical simulation of many-body fermionic systems is a notoriously difficult problem in theoretical condensed matter. In particular, quantum Monte Carlo algorithms usually fail due to the so-called sign-problem \cite{signProblem}.
Our digital quantum simulator setup enables to efficiently compute the quantum dynamics of interacting fermions, even on an arbitrary two-dimensional lattice. Although we focus on the paradigmatic Fermi-Hubbard Hamiltonian, the proposed scheme can be generalized to the quantum simulation of several other fermionic models, such as the Anderson impurity model.

\begin{figure*}[t]
   \centering
   \includegraphics[width=15cm]{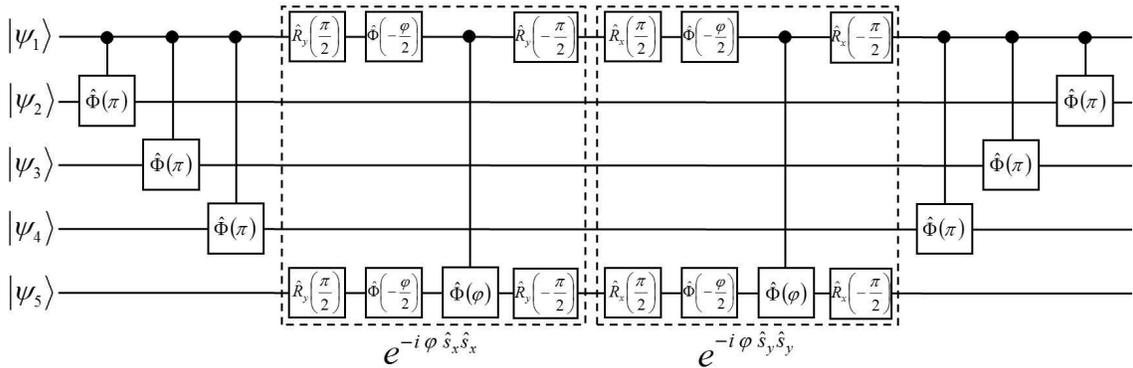}
   \caption{Quantum circuit description for the simulation of the hopping part of the Fermi-Hubbard model on a two-dimensional lattice.
   Here we explicitly show the case of $\hat{\mathcal{H}}_\lambda = -\lambda \left( \hat{c}^\dagger_1 \hat{c}_5 + \hat{c}^\dagger_5 \hat{c}_1 \right)$, with $\varphi=2 \lambda \tau$. $\hat{R}_x(\theta)$ and $\hat{R}_y(\theta)$ indicate single-qubit rotations about $x$ or $y$ axis of the Bloch sphere, while $\hat{\Phi} (\varphi)$ is the single-qubit phase gate.}
   \label{gateline}
\end{figure*}
The target Hamiltonian describing a two-dimensional $N \times M$ lattice of Wannier orbitals is
\begin{equation}
\hat{\mathcal{H}}_{Hub} = - \lambda \sum_{\langle \mu,\nu \rangle, \sigma} \hat{c}^\dagger_{\mu,\sigma} \hat{c}_{\nu,\sigma} +
                      U \sum_\mu \hat{c}^\dagger_{\mu, \uparrow} \hat{c}_{\mu, \uparrow} \hat{c}^\dagger_{\mu, \downarrow} \hat{c}_{\mu, \downarrow},
\label{Hubbard}
\end{equation}
where $\langle \mu,\nu \rangle$ are nearest neighbors ($\nu=\mu \pm 1$, $\nu=\mu \pm M$) and
$\hat{c}_{\mu,\sigma}$ are fermionic operators.
In order to simulate this Hamiltonian with our setup, we exploit the Jordan-Wigner transformation to map fermion operators $\hat{c}_\mu$ onto spin ones $\hat{s}_\mu$ \cite{JordanExt1,JordanExt2,JordanExt3}. However if such a transformation is applied to the Hubbard model (\ref{Hubbard}) in more than one dimension, the hopping (first) term results into XY spin couplings whose sign depends on the parity of the number of occupied states that are between $\mu$ and $\nu$ in the chosen ordering of the Wannier orbitals \cite{LLoydfermioni}. This aspect makes the simulation of a fermionic system much more demanding than any typical spin system, because the resulting effective spin Hamiltonian contains many-spin terms. To illustrate how we address this key issue, here we consider the simpler case of the hopping of spinless fermions on a lattice (the general case of interacting spin fermions is discussed in App.~D). The target Hamiltonian can be mapped into the following spin model:
\begin{equation}
\hat{\mathcal{H}}_\lambda = -\lambda \sum_{\langle \mu < \nu \rangle} (-1)^{\hat{\alpha}} \hat{s}^+_\mu \hat{s}^-_\nu + \text{h.c.},
\label{Ht}
\end{equation}
where $\hat{\alpha}={\sum_{\gamma=\mu+1}^{\nu-1} \hat{c}_\gamma^\dagger \hat{c}_\gamma} \equiv \sum_{\gamma=\mu+1}^{\nu-1} (\hat{s}^z_\gamma+\frac{1}{2})$.
We simulate this $n$-body interaction by taking care of the state-dependent phase, similarly to Refs.~\onlinecite{Laflamme1,Laflamme2}.
The sign factor in (\ref{Ht}) is obtained by performing a conditional evolution of the qubits interposed between the specifically addressed sites, $\mu$ and $\nu$, depending on the state of $\mu$. This corresponds to a series of controlled-Z (CZ) gates between qubit $\mu$ and each of the qubits $\gamma$ interposed between $\mu$ and $\nu$. Hence, the sequence of gates to be implemented at each Trotter step is the following:
\begin{equation}
\prod_{\mu<\gamma<\nu} \hat{U}_{CZ_{\mu,\gamma}} e^{-i \lambda \tau (\hat{s}^+_\mu \hat{s}^-_\nu + \hat{s}^+_\nu \hat{s}^-_\mu )} \prod_{\mu<\gamma<\nu} \hat{U}_{CZ_{\mu,\gamma}}.
\label{sequence}
\end{equation}
For instance, in Fig. \ref{gateline} we show the quantum circuit for the implementation of $\hat{\mathcal{H}}_\lambda^{1,5} = -\lambda (\hat{c}^\dagger_1 \hat{c}_5 + \hat{c}^\dagger_5 \hat{c}_1)$: controlled-phase gates (with $\varphi = \pi$) between qubit $|\psi_1\rangle$ and each of the qubits interposed between $|\psi_1\rangle$ and $|\psi_5\rangle$, namely $|\psi_2\rangle$, $|\psi_3\rangle$ and $|\psi_4\rangle$, are sequentially performed before and after the central block (dashed boxes), which implements the XY evolution: $\hat{U}_{XY} = \exp\{-i \varphi (\hat{s}_{1x} \hat{s}_{5x} + \hat{s}_{1y} \hat{s}_{5y}) \}$.
The latter consists of two controlled-$\varphi$ gates (with $\varphi=2 \lambda \tau$), preceded and followed by proper single-qubit rotations, implementing respectively $\hat{s}_{x} \hat{s}_{x}$ and $\hat{s}_{y} \hat{s}_{y}$ terms of the interaction, as schematically explained in Fig.~\ref{gateline}.
By exploiting the high mobility of the photons entering into the hybrid encoding, Hamiltonian terms involving distant qubits can be simulated straightforwardly.
In fact, this is a remarkable advantage with respect to alternative solid-state platforms for quantum information processing.
We stress that, in spite of the increment in the number of gates required to address the sign issue, a large number of hopping terms can still be implemented in parallel.

\section{Quantum simulations with inhomogenously broadened spin ensembles}

In this Section we show how the present scheme can be made robust against inhomogenous broadening (IB) of the SE, which is probably the major shortcoming in its use to encode quantum information. A certain degree spin  inhomogeneity is unavoidable in real SEs, and may result from slightly disordered spin environments or by random magnetic fields produced by surrounding nuclear magnetic moments. Due to IB, the collective (''super-radiant") spin-excitation that couples to the photon field spontaneously decays into a quasi-continuum of decoupled, ``dark'' spin modes within a timescale of order $\hbar /\sigma$, $\sigma$ being the width of the distribution of gaps in the SE. A possible way to deal with IB is to revert the associated Hamiltonian evolution by echo techniques, e.g. by using external magnetic field pulses resonant with the spin gaps \cite{EchoMolmer,EchoBertet}. While possible in principle, implementing such a solution within our simulation scheme would be very demanding. Pulses should act with very high fidelity, and should be controlled independently for each logical resonator with the proper timing to restore qubits before they undergo gates.
Here we prefer to adopt a different strategy, which does not require additional resources and works well with the present scheme. The idea is to run the simulation by keeping the SE in a ``cavity protection'' regime \cite{Auffeves,Molmer,cavityProtection}.
Indeed, a strong spin-resonator coupling provides a protection mechanism, by inducing an energy gap between the computational (super-radiant) and the non-computational (dark) modes \cite{Molmer}, thus effectively decoupling these two.
This mechanism has been experimentally demonstrated in Ref.~\onlinecite{cavityProtection}.\\
In the non-resonant (dispersive) regime, the energy shift of the super-radiant mode is of order $\bar{G}^2/\Delta$, where $\bar{G}$ is the collective SE-cavity coupling and $\Delta=\omega_{-1}-\omega_c(0)$ is the detuning between the resonator frequency and the spin gap.
By assuming a spin ensemble with gaussian broadening and standard deviation $\sigma$, the cavity protection condition is fulfilled if $\bar{G}^2/\Delta \gg \sigma$. However, reducing the detuning (or, equivalently, increasing the coupling strength) leads to unwanted oscillations of a significant fraction ($\sim \bar{G}/\Delta$) of the wave-function between logical states $|0\rangle$ and $|1\rangle$. A trade-off can be found in the limit of very large $\bar{G}$ (200-300 MHz) and $\Delta$ ($\sim 15-20$ GHz), but this is not within reach of present technology. Nevertheless, we show that these oscillations do not prevent to implement the proposed digital simulation scheme, thus making it possible to employ experimentally available conditions. In the following we use small detunings, i.e. $\Delta = 6 \bar{G}$, with $\bar{G} = 30$ MHz, and a SE with gaussian broadening and FWHM $=1$ MHz, as assumed in \cite{Auffeves}.
We numerically determine the time evolution of the qubit wave-function, coupled to a bath of dark modes, by exploiting the formalism developed in \cite{Auffeves}. We obtain a wave-function leakage to dark states at long times of only $\sim 1 \%$. As expected, this is lower than the upper bound ($4 \sigma^2 \Delta^2 / G^4 \approx 2.8 \%$) obtained in \cite{Molmer}. As we explicitly show below, the wave-function oscillations associated with the relatively large value of $\bar{G}/\Delta$ can be easily adjusted in our scheme, since they are coherent and correspond to single-qubit rotations.\\
Here we proceed to illustrate the use of our simulation scheme in a cavity-protected regime and specific SE parameters.
The best spin systems are the so-called $S$-ions (such as Fe$^{3+}$ or Gd$^{3+}$) whose orbital angular momentum vanishes because of Hund's rules. This makes the ion practically insensitive to disorder in the environment. In addition, the number of nuclear spins should be minimized, since these produce random quasi-static magnetic fields causing IB. Linewidths as small as a fraction of MHz are indeed observed in diluted magnetic semiconductors, such as Fe$^{3+}$ in ZnS \cite{ZnS}, whose nuclei are mostly spinless. Even if Fe$^{3+}$ and Gd$^{3+}$  are $S>1$ spins, these behave as effective $S=1$ SEs if all magnetic-dipole transitions from the ground stated are far detuned from resonator frequencies, apart from two.\\
To keep the experimental demonstration of the proposed scheme as easy as possible, we choose resonator frequencies smaller than in Section II. In particular, we assume 14 GHz for the {\it logical}, and 10.2 GHz for the {\it auxiliary} resonators, respectively, which are comparable with the resonator frequencies already employed in Ref.~\onlinecite{Schuster}. Spin ensembles fitting our scheme with a 14 GHz superconducting resonator can be easily found. For instance, Fe$^{3+}$ impurities in the same Al$_2$O$_3$ matrix employed in \cite{Schuster} display suitable gaps
with an applied magnetic field of $\sim 70$ mT forming an angle of $\sim 70^\circ$ with the anisotropy axis (given an easy plane axial anisotropy, with $D = 5.15 $ GHz \cite{spin}). As mentioned above, the relatively small detuning $\Delta = 6 \bar{G}$ needed for cavity protection induces an unwanted one-qubit oscillation with frequency $\nu=\sqrt{\bar{G}^2+\Delta^2/4}$ ($\sim 95$ MHz), which must be taken into account in our simulation scheme. This can be done in the implementation of one-qubit gates by choosing a starting time of the gate $t_i = 2 n \pi/\nu$ and by making the detuning $\Delta \gg \bar{G}$ for the short time ($\alt 1$ ns) needed to implement the phase gate, Eq. \ref{Rz}. We stress that this time is much shorter than that characterizing the damping due to IB. As far as two-qubit gates are concerned, the only part which is affected by the unwanted oscillations is the photon hopping process between {\it logical} and {\it auxiliary} cavities, whose starting time needs to be chosen again as $t_i = 2 n \pi/\nu$ (note that the {\it auxiliary} resonator does not contain a SE). In some situations (such as the simulation of $\mathcal{H}_{xz}$ terms), when a pair of qubits is subject to a different or differently ordered sequence of operations, we need to increase the detuning for one of the two qubits to freeze its evolution, while waiting the other to complete an oscillation ("rephasing").\\
To test the performance of this setup we first numerically determine the fidelity of simulations of the elementary terms in a one- and two-qubit Hamiltonian, including also decoherence effects (by solving the Liouville-von Neumann equations of motion). Table \ref{Hamunits} shows that these fidelities remain high, thus demonstrating the effectiveness of our scheme even in a cavity-protected regime. An interesting proof-of-principle experiment that could be readily performed with existing setups is the simulation of the dynamics resulting from an $XY$ interaction between two spins $s=1/2$. This is also the central step in the simulation of hopping processes in fermion Hamiltonians (dashed boxes in Fig. \ref{gateline}).
\begin{figure}
   \centering
      \includegraphics[width=8.5cm]{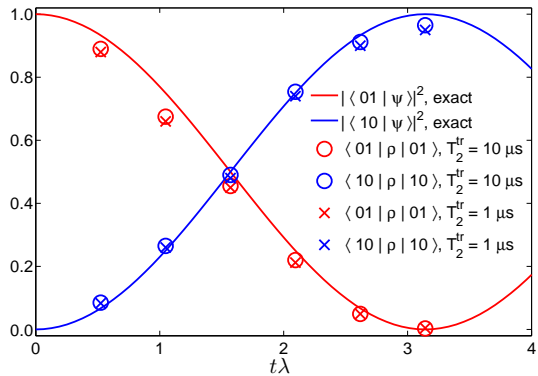}
   \caption{Simulation of the XY model on a pair of hybrid qubits using the cavity-protected regime described in the text ($\hat{\mathcal{H}}_{XY}=\lambda [\hat{s}_{1x} \hat{s}_{2x}+\hat{s}_{1y} \hat{s}_{2y}]$). Lines represent the exact evolution, whereas points are calculated with Lindblad formalism assuming $Q = 10^6$ and for two different values of $T_2^{tr}$. We use the parameters $\omega_c(0)/2 \pi=14$ GHz, $\omega_{-1}/2 \pi = 14.18$ GHz, $\omega_{1}/2 \pi = 12$ GHz, $\tilde{\omega}_c(0)/2 \pi = 10.2$ GHz, $\Omega_{01}/2 \pi = 9.2$ GHz, $\Omega_{12}/2 \pi = 8.3$ GHz.
   $\bar{G}_{-1} = 30$ MHz, $\bar{G}_{+1} = 33$ MHz, $\Delta = 6 \bar{G}$.}
   \label{evoXY}
\end{figure}
Fig. \ref{evoXY} shows that the time evolution is well reproduced even with a short $T_2^{\rm tr} =1 \; \mu$s.

\section{Discussion and perspectives}

\begin{figure}
  \centering
   \includegraphics[width=8.5cm]{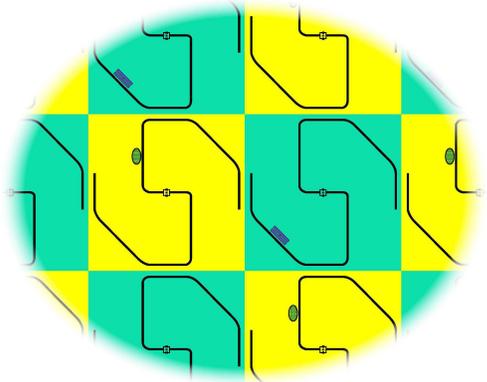}
   \caption{(Color online). Schematic representation of a two-dimensional implementation of the digital quantum simulator. Dark lines show superconducting coplanar resonators routed such that each resonator is coupled to four adjacent resonators.  Yellow boxes indicate {\it logical} resonators containing ensembles of S=1 spins near the magnetic field antinodes, while green boxes indicate {\it auxiliary} resonators containing transmons near voltage antinodes. Flux biasing of the resonator SQUIDs could be accomplished using microwave lines placed on another layer.   }
   \label{scheme2D}
\end{figure}

We have proposed a digital quantum simulator based on hybrid spin-photon qubits, encoded in an array of superconducting resonators strongly coupled to spin ensembles.
Within this quantum computing architecture, quantum gates are implemented by a single operational tool, namely by tuning the resonators frequencies. We have shown the feasibility of the scheme with state-of the-art superconducting arrays technology, which allows the high fidelity simulation of a large class of multi-qubits spin and fermionic models.
To test our predictions, we have performed numerical simulations of the master equation for the system density matrix, including the most important decoherence channels such as photon loss and pure-dephasing of the transmon involved in two-qubit entangling gates.
Coherence times of single spins are so long that their effect on quantum simulations can be disregarded. However, inhomogeneous broadening of the SE might result in an irreversible population leakage out of the computational basis. Nevertheless, we have shown in the previous Section that this problem can be circumvented by implementing the scheme in a cavity-protected regime.\\
\textit{Sources of errors}. We analyze here the sources of error that affect the quantum simulation, and point out possible solutions.
Three main simulation errors can be found: digital errors (arising from the Trotter-Suzuki approximation), gating errors (due to imperfect implementation of the desired unitaries), and decoherence errors (due to the interaction of the quantum simulator with the environment).
While digital errors can obviously be reduced by increasing the number of Trotter steps,
gating errors are accumulated by repeating a large number of quantum operations. In the same way the effects of the interaction of the system with the environment become much more pronounced if the number of gates, and thus the overall simulation time, increases.
First, we notice that the present setup {\it limits the role of the transmon}, which is not involved in the definition of the qubits. All transmons are kept in their ground states apart for the specific transmons involved in two-qubit gates, which are excited only for a short time.
Thus, typical state-of-the-art technology, which ensures transmon dephasing times of the order of tens of microseconds, is sufficient to obtain high fidelity quantum simulations of relatively large systems. Indeed, the three-spins TIM model reported here can immediately be extended to simulate spin Hamiltonians involving many spins, by addressing the different cavities in parallel.
Photon loss represents the main source of decoherence in our hybrid dual-rail encoding.
Finally, gating errors are mainly due to the small difference between the photon frequency and transmonic gaps in the auxiliary cavities, which induces a residual interaction that is never completely switched off. Here we use the tunability of the resonator frequency as the only tool to process quantum information, but the flux control of the Josephson energy of the transmons can also be exploited to increase the detuning, thus leading to even larger fidelities.\\
\textit{Two-dimensional arrays}. While any model can be implemented onto a one-dimensional register (e.g., the one schematically illustrated in Fig.~\ref{Setup}) at the cost of requiring long-range two-qubit gates, it is clear that a register topology directly mimicking the target Hamiltonian would greatly reduce the simulation effort.
In particular, there are several important Hamiltonians defined on two-dimensional lattices whose simulation would greatly benefit from a two-dimensional register. Here, we point out that our scheme is straightforwardly usable on such a register, but its experimental realization necessarily requires the implementation of two sub-lattices of cavities, alternatively coupled to spin and transmon qubits, respectively.
Fortunately, resonator arrays with complex network topologies are realistically possible, already, as each cavity can easily couple to multiple other resonators.  Fig.~\ref{scheme2D} displays the schematic drawing of a potential two-dimensional layout showing how such sub-lattices could feasibly realize a two-dimensional simulator.
>From a technological point of view, we notice that similar lattices with transmon qubits have been fabricated with more than 200 coupled cavities. While local tuning in such a lattice would require local flux bias on a separate layer, this need for local control lines applies to any adjustable quantum simulator. On the other hand, we notice that a recent technology has shown promising results to bring flux lines to the interior part of a lattice made of a small number of nodes, e.g. by using Aluminum airbridge crossovers to route microwave signals into a target resonator \cite{Martinis2014}.\\
 \textit{Summary}. In conclusion, the proposed setup exploits the best characteristics of distinct physical systems: the long coherence times of the spins, which can encode quantum information and protect it from decoherence, 
and the mobility of photons entering this hybrid encoding of qubits.
In the end, this allows to realize long-range two-body interactions between distant qubits without the need for much more demanding SWAP gates.
Moreover, on-site tunability and scalability make this architecture extremely appealing and competitive with respect to alternative proposals, either based on superconducting arrays or on different technologies.


\acknowledgments{
Very useful discussions with G. Amoretti and H. E. T\"ureci are gratefully acknowledged. This work has been financially supported by FIRB Project No. RBFR12RPD1 of the Italian Ministry of Education and Research (MIUR)}.

\appendix

\section{Hybrid dual-rail encoding}
We consider a coplanar waveguide resonator containing a single photon in a mode of frequency $\omega_c$, and an ensemble of $\mathcal{N}$ non-interacting and equally oriented $s=1$ spins. In the low-excitation regime, the SE can be modeled by two independent harmonic oscillators, related to two different magnetic-dipole transitions from the
$m = 0$ ground state of the single spin, to the $m = - 1$ and $m = 1$ states, with excitation frequencies $\omega_{-1}$ and $\omega_{1}$. This can be achieved by properly choosing a system with easy-plane magnetic anisotropy, which provides a zero-field splitting between the $m = 0$ ground state and the excited $m = \pm 1$ doublet, and in the presence of a small static magnetic field.
We suppose to initialize the system by preparing each spin in its ground state:
$\left|\phi_0\right\rangle \equiv |0_1 ...0_\mathcal{N} \rangle$. \\
If the resonator frequency is tuned to match the spin gap $\omega_{1}$, the SE can absorb the photon and collectively evolve into the state $\left|\phi_1\right\rangle = \frac{1}{\sqrt{\mathcal{N}}} \sum_{q=1}^{\mathcal{N}} \left| 0_1 ... 1_q ... 0_{\mathcal{N}} \right\rangle$. Transitions between $ \left|\phi_0\right\rangle $ and $ \left|\phi_1\right\rangle $ are described (in the limit of low number of excitations) by the bosonic operators $\hat{b}_1$ and $\hat{b}_1^{\dagger}$, where $\hat{b}_1=\frac{1}{\sqrt{\mathcal{N}}} \sum_{q=1}^\mathcal{N} |  0 \rangle\langle 1  |_q $ and $[\hat{b}_1,\hat{b}_1^\dagger] = 1$ \cite{Wesenberg,KuboPRL}. Conversely, if the resonator frequency is tuned to $\omega_{-1}$, the SE can evolve into the state $\left|\phi_{-1}\right\rangle = \frac{1}{\sqrt{\mathcal{N}}} \sum_{q=1}^{\mathcal{N}} \left| 0_1 ... -1_q ... 0_{\mathcal{N}} \right\rangle$, the transition being described by the operators $\hat{b}_{-1}$ and $\hat{b}_{-1}^{\dagger}$, where $\hat{b}_{-1}=\frac{1}{\sqrt{\mathcal{N}}} \sum_{q=1}^\mathcal{N} |  0 \rangle\langle -1  |_q $. \\
Within the single-excitation subspace of the system formed by the cavity mode and the SE, we introduce the hybrid dual-rail encoding of the qubit $\mu$:
\begin{eqnarray}\label{encoding}\nonumber
| 0 \rangle_\mu &\!\equiv\!& \hat{b}_{-1,\mu}^\dagger | \emptyset \rangle \!=\! |\phi_{-1}^\mu, n_\mu\!=\!0 \rangle , \\
| 1 \rangle_\mu &\!\equiv\!& \hat{a}_\mu^\dagger | \emptyset \rangle \!=\! |\phi_0^\mu, n_\mu\!=\!1 \rangle ,
\end{eqnarray}
where $ \hat{a}_\mu^\dagger $ is the photon creation operator and
$| \emptyset \rangle = | \phi_0 , n_\mu = 0 \rangle $ is the vacuum state.

\section{Single- and two-qubit gates}
\subsection{Single-qubit rotations}
Resonant processes involving the absorption (emission) of the photons entering the hybrid encoding in (Eq. \ref{encoding}) are exploited to perform one- and two-qubit gates.
These processes are induced by ``shift pulses'', in which the frequency of cavity $\mu$ is varied by a quantity $\delta_c^\mu$ for a suitable amount of time.
In the idle mode, the photon frequencies are largely detuned from the spin energy gaps, and $\hat{H}_{int}$ is ineffective. In addition, the modes $\omega_c^\mu$ and $\tilde{\omega}_c^j$ of neighboring cavities are far-detuned and the effect of $\hat{H}_{ph-ph}$ is negligible. Single-qubit gates can thus be performed independently on each qubit, which can be individually addressed.\\
Off-resonance pulses are employed to obtain a rotation by an arbitrary angle about the  $z$ axis of the Bloch sphere. These induce a phase difference between the $|0\rangle$ and $|1\rangle$ states of the hybrid qubits (Eq. \ref{encoding}) and performs the well-known phase gate:
\begin{equation}
\Phi (\delta_c^\mu T) =
\left(
\begin{array}{cc}
1 & 0 \\
0 & e^{-i \delta_c^\mu T}
\end{array}
\right) .
\label{Rz}
\end{equation}
where we have assumed step-like pulses of amplitude $\delta_c^\mu$ and duration $T$.\\
Conversely, resonant pulses are employed to transfer the excitation between SEs and resonators. This produces a generic rotation in the $x$-$y$ plane of the Bloch sphere:
\begin{equation}
  R_{xy} (\theta) \! = \! \left(\!\!
\begin{array}{cc}
\text{cos} \,( \theta/2) & -i~ e^{-i \delta_c^\mu t_0 } \, \text{sin}\, (\theta/2) \\
-i e^{i \delta_c^\mu t_0 } \,\text{sin} \,(\theta/2) & \text{cos} \,(\theta/2)	
\end{array}
\!\! \right) ,
\label{Rxy}
\end{equation}
with $\theta = \bar{G}_{-1} T$.
By properly tuning the initial time we can obtain rotations about $x$ ($\delta_c^\mu t_0 = 2 k \pi$) or $y$ ($\delta_c^\mu t_0 = (4 k+1) \pi/2$) axis, while the pulse duration controls the rotation angle.
See Ref. \cite{PRAcav} for a detailed derivation. \\

\subsection{Controlled-phase gate}
The Controlled-phase (C$\varphi$) two-qubit gate is represented by the matrix:
\begin{equation}
U_{C{\varphi}} = \left(
\begin{array}{cccc}
~1 & ~0 & ~0 & ~0 \\
~0 & ~1 & ~0 & ~0 \\
~0 & ~0 & ~1 & ~0 \\
~0 & ~0 & ~0 & e^{-i \varphi} 	
\end{array}
\right)
\, \, .
\end{equation}
It can be implemented by means of two-step semi-resonant Rabi oscillations of the transmon state between $| \psi_{0,j} \rangle$ and $| \psi_{2,j} \rangle$.
We describe here the C$\varphi$ multi-step pulse sequence on two qubits initialized in the state
$ |1_\mu 1_\nu \rangle $, as schematically shown in Fig. \ref{Setup}-(b) for $\mu=2$, $\nu=3$ and $j=2$:
\begin{enumerate}
\item The first step corresponds to the hopping of the photon from {\it logical} cavity 3 to the {\it auxiliary} resonator 2 (interposed between qubits 2 and 3), by means of a $\pi$-pulse that brings the two cavities into resonance.
\item As a second step, the frequency of resonator $\mu=2$ ($\tilde{\omega}_c^2$) is tuned to $\Omega_{01}$ by
means of a $\pi$-pulse, which transfers the excitation to the intermediate level $|\psi_{1,j=2}\rangle$ of the transmon.
\item A $\pi$-pulse is exploited to induce the hopping of a second photon from {\it logical} cavity 2 to the {\it auxiliary} resonator.
\item Then, a semi-resonant process (during which the resonator is detuned from the transmon gap by a small amount $\delta_{12}$) is exploited to induce an arbitrary phase on the $|1_2 1_3 \rangle $ component of the wavefunction \cite{Mariantoni}. A pulse of duration $\Delta t = \frac{\pi}{\sqrt{G_{12}^2+\delta_{12}^2/4}}$, where $\delta_{12} = \Omega_{12}-\tilde{\omega}_c^2$ is the detuning between the resonator mode and the $|\psi_{1,2}\rangle \rightarrow |\psi_{2,2}\rangle$ transition of the transmon, adds a phase $\varphi = \pi - \pi \frac{\delta_{12}}{\sqrt{\delta_{12}^2+ 4 G_{12}^2}}$ to the system wavefunction.
\item Finally, the repetition of the first three steps brings the state back to $ |1_2 1_3\rangle $, with an overall phase $\varphi$. By properly setting the delay between the two $\pi$ pulses corresponding to the previous steps (or by performing single-qubit phase shifts), the associated absorption and emission processes yield a zero additional phase.
\end{enumerate}
Conversely, the other basis states do not acquire any phase, as required for the C${\varphi}$ gate, due to the absence of at least one of the two photons (see Ref. \cite{PRAcav}). For $\delta_{12} = 0$ we obtain the usual full Rabi process, which implements a Controlled-Z (CZ). \\
The setup is simplified with respect to our previous proposal \cite{PRLcav}, as each resonator contains a single photonic mode.\\
It is also important to note that here we are using an ensemble of effective spins $S=1$ as this ensures the possibility of implementing Controlled-phase gates between distant qubits, with no need of performing highly demanding and error-prone sequences of two-qubit SWAP gates.
Long-distance two-qubit interactions are a key-resource for the digital simulation of many interesting physical Hamiltonians. They appear each time that a multi-dimensional target system is mapped onto a linear chain of qubits or in models with $N$-body terms. Among these, as discussed in Section III.B, a particular interest is assumed by problems involving interacting fermions in two or higher spatial dimensions, which are often intractable for classical computers.
For instance, solving the two-dimensional Hubbard model is considered by many as
the ultimate goal of the theory of strongly correlated systems.
In these cases the Jordan-Wigner mapping induces many-spin interactions \cite{Laflamme2} which can be handled as outlined in Fig. \ref{gateline}, provided the ability to efficiently implementing long-range two-qubit couplings.
These are obtained by bringing the photon components of the two qubits into neighboring {\it logical} resonators by a series of hoppings. The operations  outlined in Fig. \ref{Setup}-(b) are then performed to implement a C$\varphi$ gate between neighboring qubits, and the photon components are finally brought back to the starting position by reverting the series of hoppings.
The photons can be transferred with negligible leakage and without perturbing the interposed qubits by temporarily storing the photon component of these qubits into the $m=1$ spin oscillator.
We stress that a large number of these long-range two-qubit gates can be implemented in parallel in the actual setup.

\section{Density matrix master equation}
The time evolution of the system density matrix $\hat{\rho}$ is described within a Markovian 
approximation and a Lindblad-type dynamics, with the Liouville-von Neumann equation of motion \cite{Scully}:
\begin{equation}
\frac{d}{dt}{\hat{\rho}} = - i \left[ \hat{H}, \hat{\rho} \right] + \sum_{i}  \Gamma_i \mathcal{L}_{\hat{x}_i}[\hat{\rho}] + \sum_{i} \gamma_i \mathcal{L}_{\hat{x}_i^\dagger \hat{x}_i}[\hat{\rho}] ,
\label{evorho}
\end{equation}
being $\Gamma_j$ and $\gamma_j$ respectively the damping and pure-dephasing rates of the field $\hat{x}_j$. The Lindblad term for an arbitrary operator, $\hat{x}$, is given by
\begin{equation}
\mathcal{L}_{\hat{x}} [\hat{\rho}] = -\frac{1}{2} \left( \hat{x}^\dagger \hat{x} \hat{\rho} + \hat{\rho} \hat{x}^\dagger \hat{x} \right) + \hat{x} \hat{\rho} \hat{x}^\dagger.
\label{Lindblad}
\end{equation}
If the operator $ \hat{x}_i $ destroys an excitation in the system, terms like $\mathcal{L}_{\hat{x}_i}[\hat{\rho}]$ account for energy losses, while pure dephasing processes are described by $\mathcal{L}_{\hat{x}_i^\dagger \hat{x}_i}[\hat{\rho}]$. We note \cite{PRAcav} that the former ones provide the most important contribution for photons \cite{t2ph} (with $\hat{x}_i = \hat{a}_\mu$, $\tilde{\hat{a}}_j$), while the latter are very important for the transmons  ($\hat{x}_i = |\psi_{k,j}\rangle \langle \psi_{k+1,j}|$, $k=0,1$).
We represent each field as a matrix in the Fock-states basis, and truncate it at a number of total excitations previously checked for convergence. The total Hamiltonian, Eq. (\ref{hamil}), and the density matrix master equation of the whole system, Eq.~(\ref{evorho}), are built by tensor products of these operators. Then, the equation of motion for $\hat{\rho}$ is numerically integrated, in the interaction picture, by using a standard Runge-Kutta approximation.

\section{Interacting spin fermions}
To extend the quantum simulation of two-dimensional Hubbard models to the case of fermionic systems with spin, we need to encode each fermion operator into a pair of qubits, corresponding to spin up and spin down.
To achieve this, we exploit a generalization of the Jordan-Wigner transformation \cite{Bari}. For this mapping we need to introduce two different spin 1/2 operators $\vec{\hat{S}}$ and $\vec{\hat{T}}$, with
$\hat{S}^z_{2 \mu-1} = \hat{c}^\dagger_{\mu \uparrow} \hat{c}_{\mu \uparrow}-\frac{1}{2}$ and $\hat{T}^z_{2\mu} = \hat{c}^\dagger_{\mu \downarrow} \hat{c}_{\mu \downarrow}-\frac{1}{2}$, describing respectively odd and even qubits (ordered by rows in the two-dimensional lattice).
\begin{eqnarray}\nonumber
\hat{S}^+_{2\mu-1} = \hat{c}^\dagger_{\mu \uparrow} e^{i \pi \sum_{\nu=1}^{N M} \hat{c}^\dagger_{\nu \downarrow} \hat{c}_{\nu \downarrow} + i \pi \sum_{\nu=1}^{\mu-1} \hat{c}^\dagger_{\nu \uparrow}\hat{c}_{\nu \uparrow} } \\
\hat{T}^+_{2\mu} = \hat{c}^\dagger_{\mu \downarrow} e^{i \pi \sum_{\nu=1}^{\mu-1} \hat{c}^\dagger_{\nu \downarrow} \hat{c}_{\nu \downarrow}}.
\label{spinmap}
\end{eqnarray}
It can be shown that these operators satisfy the usual angular momentum commutator algebra, and that $\left[ \hat{S}^\alpha_{2\mu-1}, \hat{T}^\alpha_{2\nu} \right] = 0$. We assume that the fermion variables are ordered by rows in the Hamiltonian. The efficiency of the scheme would be increased by using a 2-dimensional setup consisting of $N$ rows and $2 M$ columns. We can write the Hubbard Hamiltonian
in terms of the spin variables introduced above
\begin{eqnarray}\nonumber
\hat{H}_{Hub}^s \!\!\! &=& \!\!\! -\lambda \sum_{\mu,\nu>\mu=1}^{N M} \!\!\! (-1)^{\sum_{\gamma=\mu+1}^{\nu-1} (\hat{S}^z_{2 \gamma-1}+\frac{1}{2})} \hat{S}^+_{2\mu-1} \hat{S}^-_{2\nu-1} + \text{h.c.} \\
&-& \lambda \sum_{\mu,\nu>\mu=1}^{N M} \!\!\! (-1)^{\sum_{\gamma=\mu+1}^{\nu-1} (\hat{T}^z_{2 \gamma}+\frac{1}{2})} \hat{T}^+_{2\mu} \hat{T}^-_{2\nu} + \text{h.c.} \\ \nonumber
&+& \!\!\! U \sum_{\mu=1}^{N M} \hat{S}^z_{2\mu-1} \hat{T}^z_{2\mu} + \! \frac{U}{2} \sum_{\mu=1}^{N M} \left(  \hat{S}^z_{2\mu-1} \! + \hat{T}^z_{2\mu} \right) \! + \! \frac{N M U}{4} ,
\label{Hubbardspin}
\end{eqnarray}
where $\mu$ and $\nu$ are nearest neighbors on the two-dimensional fermionic lattice, such that $\nu=\mu + 1$ (horizontal neighbors) or $\nu= \mu + M$ (vertical neighbors) with the present labeling. Odd (even) qubits encode spin up (spin down) variables.
Since the hopping term does not act if $\langle \hat{c}^\dagger_{\mu \sigma} \hat{c}_{\mu \sigma} \rangle = 1$ (i.e. $\hat{c}^{\dagger2}_{\mu\sigma} = 0$), we can start directly with $\gamma=\mu+1$, and the exponential in expressions like $\hat{S}^+_\mu \exp\{i \pi \sum_{\gamma=\mu+1}^{\nu-1} (\hat{S}^z_{\gamma } +\frac{1}{2})\} \hat{S}^-_\nu$ can be factorized.
We note that in the case of horizontal neighbors the phase factor cancels out and that in $\hat{H}_{Hub}^s$ do not appear terms $\hat{S}^+_{2\mu-1} \hat{T}^-_{2\nu}$, as we are not considering spin-flip processes.\\
To simulate such evolution we can proceed in a way analogous to the spinless case. Here, however, two different series of $CZ_{\mu,\gamma}$ should be carried out, depending if we are considering the hopping of spin $\uparrow$ or spin $\downarrow$ fermions. The former involves only odd values of $\gamma$, the second only even.
Notice that, in a 2-dimensional register, we need to transfer photons to implement $\hat{S}^+_{2\mu-1} \hat{S}^-_{2\nu-1}$ or $\hat{T}^+_{2\mu} \hat{T}^-_{2\nu}$ each time we have to couple a pair of fermions belonging to the same row (due to the alternating $\uparrow$-$\downarrow$ mapping), but in that case $\prod \hat{U}_{CZ_{\mu,\gamma}}$ is not required. The term $\prod \hat{U}_{CZ_{\mu,\gamma}}$, needed to correct the {\it sign problem}, is necessary only if $\nu = \mu + M $ (no photon transfer in that case is needed).

\end{document}